\documentclass{article}%
\usepackage{amsfonts}
\usepackage{amsmath}
\usepackage{amssymb}
\usepackage{graphicx}%
\providecommand{\U}[1]{\protect\rule{.1in}{.1in}}

\begin{document}

\title{The Role of Correlation in Quantum and Classical Games}
\author{\textit{Simon J.D. Phoenix \& Faisal Shah Khan}\\Khalifa University, Abu Dhabi, PO Box 127788, UAE}
\maketitle

\begin{abstract}
We use the example of playing a 2-player game with entangled quantum objects
to investigate the effect of quantum correlation. We find that for simple game
scenarios it is \textit{classical} correlation that is the central feature and
that these simple quantum games are not sensitive to the quantum part of the
correlation. In these games played with quantum objects it is possible to
transform a game such as Prisoner's Dilemma into the game of Chicken. We show
that this behaviour, and the associated enhanced equilibrium payoff over
playing the game with quantum objects in non-entangled states, is entirely due
to the classical part of the correlation.

Generalizing these games to the pure strategy 2-player quantum game where the
players have finite strategy sets and a projective joint measurement is made
on the output state produced by the players, we show that a given quantum game
of this form can always be reproduced by a classical model, such as a
communication channel. Where entanglement is a feature of the these 2-player
quantum games the matrix of expected outcomes for the players can be
reproduced by a classical channel with correlated noise.

\end{abstract}

\section{Introduction}

The field of computer science has been revolutionised by the realisation that
computers are physical objects obeying physical laws. Allowing computational
devices to access the features of quantum mechanics, and entanglement in
particular, has resulted in the potential for quantum devices that can perform
certain computations significantly faster than their classical counterparts,
lowering the complexity class of the associated problems. It was an innovative
and groundbreaking step to ask the same question of classical game theory.
Could the introduction of quantum objects and operations to the theory of
games also result in a similar revolution allowing resolutions and
enhancements not available within a classical treatment?

Since the seminal work of Eisert, Wilkens and Lewenstein (EWL) and Meyer [1,2]
quantum games have been the subject of much work and controversy. The question
of whether a game played with quantum objects can be considered to be quantum
mechanical at all has been raised (see, in particular [3]) and the necessity
of comparing like with like within the context of games has been beautifully
formulated by Bleiler [4]. Much of the work has focused, naturally, on the use
of entangled states (see [4-21] for a small selection of the extensive
literature). In this paper we focus on 2-player non-cooperative games in which
a single projective quantum measurement is performed to generate the
measurement results over which the players have preferences. We address the
question of whether such games can access the \textit{quantum} nature of any
correlation and also question to what extent such games can be considered to
be truly quantum mechanical, requiring quantum objects in order to achieve a
given result or outcome for the players.

We begin by considering some very simple, and restricted, examples of games
played with quantum objects in order to gain an insight into the role of
correlation in these systems. It is shown that, for these games, the results
depend only on the classical component of the correlation. For these games,
although we may begin with preferences over measurement results that are those
of one game form, the actual game the players play is a different game
altogether. In our first simple example we examine a game played with quantum
objects that would appear to be a version of Prisoner's Dilemma in the first
instance. Upon closer inspection, however, it can be seen that the players are
\textit{actually} playing the classical game of Chicken\footnote{Of course,
`changing' a game by considering an appropriate extension of it is nothing new
within game theory. The physics of playable games [13] is highlighting this
here in a rather dramatic fashion.} (for an excellent text on game theory see,
for example, [22]).

We show that the same game transformation can be achieved by a classical game
in which the players' choices are communicated over a noisy channel with a
classical correlated noise process. Thus the properties of the quantum game
are not dependent upon the quantum part of the correlation. By extending the
classical game to the mixed case we see that a classical correlated noise
process can lead to similar enhancements of equilibrium payoffs as that
claimed for the quantum game with a full quantum strategy set. The role of the
classical correlations is further highlighted by consideration of a simple
quantum game in which the players communicate their choice by the transmission
of their respective particles over some noisy channel such that the quantum
interference terms in the density matrix are suppressed. In this scenario we
see that an enhanced equilibrium and the transformation of the game is also
obtained. These results strongly suggest that the enhancements and the game
transformation are due to \textit{classical} correlations in these 2-player
games rather than any specifically quantum mechanical feature of the correlation.

We develop a general approach to simple 2-player games played with quantum
objects that allows the analysis of a wider class of games than our initial
examples. Thus we adopt the perspective that games can be played with quantum
objects (we \textit{game} the quantum [23-25]) rather than worry about whether
such games are proper extensions of some underlying classical game. We show
that such games can always be thought of as being \textit{equivalent} to a
classical game played with classical coins in the sense that the players
analyse the game as if they were playing the equivalent classical game with a
potentially different set of preferences to those of the initial quantum game.
In this way we can see that a given classical game may sometimes be thought of
as a decomposition consisting of a different quantum game with different
preferences. Our simple example shows that 2 players can play the game of
Chicken by playing a version of Quantum Prisoner's Dilemma [23].

By focusing on the notion of \textit{playable} games, that is there is an
implementation of the game with physical objects, we describe the general
features of any game whether played with classical or quantum objects. The
requirement of playability allows us to develop a model in which the elements
necessary for proper comparison of quantum vs classical behaviour are made
clear. We believe that this approach, grounded in the physics of the game
objects and mechanisms, gives a perspective on quantum games that helps to
clarify the issue of just what is quantum mechanical in a quantum game.

\section{Turning Prisoners into Chickens}

Let us consider an attempt to implement the game of Prisoner's Dilemma (PD)
using quantum objects and operations. We shall assume the players (Alice and
Bob) are each given a spin-1/2 particle upon which to operate. The players are
each allowed only two operations on their own particles; flip or don't-flip,
with respect to the spin-$z$ direction. We shall label these operations as
\textit{F} and \textit{I}, respectively. There will be, in general, 4 possible
output states that the players can produce, characterized by the choices
$\left(  I,I\right)  ,\left(  I,F\right)  ,\left(  F,I\right)  $ and $\left(
F,F\right)  $ where the first element refers to the choice of Alice and the
second to that of Bob. The output state is subject to a measurement as
follows; the spin in the $z$-direction of Alice's particle is measured and the
spin in the $z$-direction of Bob's particle is measured. The possible
measurement results are listed as a tuple $\left(  0,0\right)  ,\left(
0,1\right)  ,\left(  1,0\right)  $ or $\left(  1,1\right)  $ where the `0'
result indicates spin-down. The measurement results are mapped to an outcome
tuple for the players so that%

\begin{align}
\left(  0,0\right)   &  \longrightarrow\left(  3,3\right) \nonumber\\
& \nonumber\\
\left(  0,1\right)   &  \longrightarrow\left(  0,5\right) \nonumber\\
& \nonumber\\
\left(  1,0\right)   &  \longrightarrow\left(  5,0\right) \nonumber\\
& \nonumber\\
\left(  1,1\right)   &  \longrightarrow\left(  1,1\right)
\end{align}
The preferences of the players are thus encapsulated by the assignment of a
numerical value as an outcome. If the initial state of the spin-1/2 particles
is given, in the measurement basis, by $\left\vert 0\right\rangle _{A}%
\otimes\left\vert 0\right\rangle _{B}$ which we shall write as $\left\vert
00\right\rangle $ then the payoff matrix becomes

\begin{quote}%
\[%
\begin{tabular}
[c]{c|cc}%
\textit{A}%
$\backslash$%
\textit{B} & $I$ & $F$\\\hline
&  & \\
$I$ & $\left(  3,3\right)  $ & $\left(  0,5\right)  $\\
&  & \\
$F$ & $\left(  5,0\right)  $ & $\left(  1,1\right)  $%
\end{tabular}
\
\]
\textbf{Table 1}: \textit{the outcome matrix for an implementation of
classical PD using quantum objects}
\end{quote}

This is nothing more than the standard description of classical Prisoner's
Dilemma [22].

Now let us consider playing the game with a \textit{different} input state of
the particles, but keeping everything else the same. The players have the same
preferences over the measurement results. We choose an input state that is not
mapped onto an eigenstate of the measurement operator by the actions of the
players. This means that the results of the measurement will be distributed
according to some probability distribution. Let us choose the following
initial state%

\begin{equation}
\left\vert \psi_{0}\right\rangle =\sqrt{\frac{3}{5}}\left\vert 00\right\rangle
+\sqrt{\frac{2}{5}}\left\vert 11\right\rangle
\end{equation}
The four possible output states are given by%

\begin{align}
\left\vert \psi\right\rangle _{II}  &  =\sqrt{\frac{3}{5}}\left\vert
00\right\rangle +\sqrt{\frac{2}{5}}\left\vert 11\right\rangle \nonumber\\
& \nonumber\\
\left\vert \psi\right\rangle _{IF}  &  =\sqrt{\frac{3}{5}}\left\vert
01\right\rangle +\sqrt{\frac{2}{5}}\left\vert 10\right\rangle \nonumber\\
& \nonumber\\
\left\vert \psi\right\rangle _{FI}  &  =\sqrt{\frac{3}{5}}\left\vert
10\right\rangle +\sqrt{\frac{2}{5}}\left\vert 01\right\rangle \nonumber\\
& \nonumber\\
\left\vert \psi\right\rangle _{FF}  &  =\sqrt{\frac{3}{5}}\left\vert
11\right\rangle +\sqrt{\frac{2}{5}}\left\vert 00\right\rangle
\end{align}
Let us suppose the players choose the operation tuple $\left(  I,I\right)  $.
We can see that if they choose these operations then they will obtain the
output tuple $\left(  3,3\right)  $ with probability 3/5 and the output tuple
$\left(  1,1\right)  $ with probability 2/5. Thus, for this choice of
operations they will obtain an \textit{expected} outcome tuple of $\left(
11/5,11/5\right)  $. Doing a similar calculation for the other possible output
states we obtain the matrix for the expected outcome tuples as

\begin{quote}%
\[%
\begin{tabular}
[c]{c|cc}%
\textit{A}%
$\backslash$%
\textit{B} & $I$ & $F$\\\hline
&  & \\
$I$ & $\left(  11/5,11/5\right)  $ & $\left(  2,3\right)  $\\
&  & \\
$F$ & $\left(  3,2\right)  $ & $\left(  9/5,9/5\right)  $%
\end{tabular}
\ \ \
\]
\textbf{Table 2}:\textit{ the outcome matrix for a game played with quantum
objects having preferences over the measurement results in accord with
Prisoner's Dilemma, but in which the input quantum state is entangled
according to equation (2)}
\end{quote}

The players will use this new matrix to determine their choice of strategy.
Their \textit{actual} choice of play is thus determined by this matrix of
expected outcomes. However, the matrix of expected outcomes is nothing more
than a numerical encapsulation of the preferences of the game of Chicken. The
game of Chicken [22] can be described by the preferences%

\begin{align}
\text{Alice }  &  \text{: }O_{FI}>O_{II}>O_{IF}>O_{FF}\nonumber\\
& \nonumber\\
\text{Bob }  &  \text{: }O_{IF}>O_{II}>O_{FI}>O_{FF}%
\end{align}
where $O_{FI}$, for example, describes the outcome tuple when Alice plays $F$
and Bob plays $I$. In the game of Chicken the usual scenario is to imagine two
somewhat irresponsible youths hurtling towards one another in their cars. The
winner of the game is the one who doesn't swerve ($O_{FI}$ in which Alice
doesn't swerve, but Bob does, is Alice's most preferred outcome). If neither
swerve $\left(  O_{FF}\right)  $ then they crash and this is the least
preferred outcome for both players. If both swerve then they are both
`chickens', which is more preferable than crashing, but not as preferable as
winning. If Alice swerves, but Bob doesn't, then this is preferable to
crashing, but Alice has lost face (she is the chicken and Bob isn't) and so is
not as preferable to her as when they both swerve.

Thus for our quantum game with this entangled state input, despite initially
having preferences over the measurement results in accordance with those of
Prisoner's Dilemma, these are transformed into preferences over the expected
outcomes that are in accordance with the game of Chicken. The game that the
players \textit{actually} play is the classical game of Chicken, despite
setting up the game as a quantum version of Prisoner's Dilemma. The players
analyze their choices in terms of this matrix of expected outcomes. At this
point it is irrelevant to the players, as far as their game objectives are
concerned, whether this matrix has been generated by some complicated quantum
process or whether it has been generated by some game in which classical coins
are mapped to the outcomes of Chicken. This feature of game transformation has
been examined in a general case by consideration of measurement of the Schmidt
observables in entangled quantum games [10,11].

This simple example, based on the quantum game scenario of Marinatto and Weber
(MW) [21], illustrates that care must be taken in ascribing quantum behaviour
to a game scenario in which the objects used to implement a game are quantum
mechanical in nature. Whilst the game scenario of MW raises legitimate
concerns about whether it constitutes a proper extension of an underlying
classical game [4] it is, nevertheless, a perfectly acceptable example of a
game played with quantum objects. If we imagine the players are given some
black box with dials for their strategy choices, then if they know the payoff
function for the possible choices, they will base their final choice upon the
analysis of this payoff function. Their transformed preferences over these
expected payoffs \textit{define the actual game they are playing}, despite the
possibility that their initial preferences over the measurement results were
those of another game form. In the example we have discussed we can see that,
as far as the players are concerned, the same game can be implemented with
either quantum objects and operations or with classical coins. In other words
the players cannot tell whether the objects inside their black box are
classical or quantum mechanical.

\section{Playable Games and Quantum/Classical Comparisons}

Any game that is actually \textit{playable} must have an implementation in the
physical world. The strategy choices represent some manipulation of physical
entities be those entities classical coins or quantum particles. For any
playable game, quantum or classical, the following elements must be present

\begin{itemize}
\item some physical objects prepared in an initial state

\item a set of manipulations that can be performed on these objects. The
manipulations that are possible are the available strategy choices of the players

\item some measurement of the state of these objects after the manipulations
of the players have been carried out

\item a mapping of the measurement results to some outcomes over which the
players have different preferences
\end{itemize}

In a simple classical game, such as normal form Prisoner's Dilemma, the
initial state and measurement elements are implicit since there is a
one-to-one correspondence between the elements, and the strategy choices can
be directly related to the measurement results. If we wish to play a game
using quantum objects and operations these elements must be made explicit and
there is no longer a direct one-to-one correspondence between the strategy
choices and the measurement results, in general.

This general description of any playable game, quantum or classical, allows us
to make correct comparisons between different game versions. For example, in
our simple example of the previous section we assume an entangled input state
of the particles. The players can perform local operations on their respective
particles, which in this example is restricted to just flip or no-flip.
However, the operations performed by the players affect the \textit{entire}
quantum state which must be considered to be a single entity. Thus the correct
quantum classical comparison to draw is between the quantum game and classical
games in which the players have some ability to affect each other's coins in
some way. As we shall see, this can be achieved by the simple expedient of
assuming a \textit{classical correlated noise} on the communication of the
strategy choices of the players to the device that measures this communication
and assigns the payoffs.

Similarly, in comparing classical and quantum games we must, as a minimum
condition, give the players strategy sets of the same size. It makes little
sense, for example, to compare a classical game in which the players have only
2 strategy choices each with a quantum game in which the players have 4
available strategies each. At best we could describe this as a possible
quantum extension of the classical game, but then we must compare the quantum
extension with the relevant classical extension of the game in order to draw a
direct comparison [4]. Thus, if we are to consider entangled quantum games we
must compare these to classical games in which correlation features in some
way, otherwise the comparison is essentially meaningless. In other words we
must compare the relevant extensions of the game in the classical and quantum domains.

Let us consider 2-player games in which the players, Alice and Bob, have the
respective available operations $\left\{  \alpha_{1},\alpha_{2},\ldots
,\alpha_{p}\right\}  $ and $\left\{  \beta_{1},\beta_{2},\ldots,\beta
_{q}\right\}  $. We shall assume that the physical objects to be manipulated
are prepared in some initial state $\psi_{in}$. The output is therefore a
state of the form $\psi_{out}=\beta_{k}\alpha_{j}\psi_{in}$ where this
description is applicable to both quantum and classical game scenarios. This
operational perspective also highlights the possibility that certain games can
be non-commutative so that the order of play matters [23]. There is then some
measurement on the output state which yields a set of possible measurement
results. These measurement results are the input to the payoff function.

It is customary in entangled quantum games to consider that the players
perform independent local operations on their respective particles and to
identify a strategy choice with a quantum spin state (or the associated
unitary operator that generates this from some specified initial state). This,
however, is something of an illusion. If we consider a game of the form
considered by EWL [1] then the output state is given by%
\begin{equation}
\psi_{out}=E^{-1}\beta_{k}\alpha_{j}E\psi_{in}=E^{-1}\beta_{k}EE^{-1}%
\alpha_{j}E\psi_{in}%
\end{equation}
Here we note that $E$ acts on the entire input state and should not be
confused with the local operations that are usually assumed in treatments of
quantum games. When $E$ is the entanglement operator that produces a maximally
entangled state from the `ground' state of two spin-1/2 particles then the
strategy sets of the players are equivalent to the sets%
\begin{equation}
\left\{  \tilde{\alpha}_{1},\tilde{\alpha}_{2},\ldots,\tilde{\alpha}%
_{p}\right\}  ~~\text{and \ }\left\{  \tilde{\beta}_{1},\tilde{\beta}%
_{2},\ldots,\tilde{\beta}_{q}\right\}
\end{equation}
where%
\begin{equation}
\tilde{\alpha}_{j}=E^{-1}\alpha_{j}E~~\text{and \ }\tilde{\beta}_{k}%
=E^{-1}\beta_{k}E
\end{equation}
and so the possible manipulations of the players involve directly interacting
with the spin of their opponent, even if the sets $\left\{  \alpha_{1}%
,\alpha_{2},\ldots,\alpha_{p}\right\}  $ and $\left\{  \beta_{1},\beta
_{2},\ldots,\beta_{q}\right\}  $ represent strictly local operations. The
identification of a so-called quantum strategy with a spin state of a single
particle is, therefore, nothing more than a convenient illusion for entangled
quantum games; a game involving entangled states is formally equivalent to a
game in which we allow the players entanglement operations as part of their
strategy sets.

If we are to make a sensible comparison between quantum and classical games in
an attempt to elucidate genuine quantum behaviour we must compare like with
like. For games of the EWL type where $E$ is an entanglement operator,
therefore, we must compare with the extension of the classical game to include
correlation. It is critical, therefore, that the role of correlation is
understood in both classical and quantum games. In the following simple game
examples we demonstrate that features that may be initially considered to
arise from the quantum-mechanical nature of a correlation actually arise from
only the classical component of the correlation in an entangled state.

\section{Preservation of Preferences}

The simple example discussed above shows that our original preferences (over
the individual measurement results) in a quantum game may be transformed by
the measurement process into different preferences over the expected outcomes.
It is the probabilistic mapping induced, in general, by the quantum
measurement that forces the final analysis of the game in terms of a matrix of
expected outcomes, and these expected outcomes can generate tuples that do not
correspond to the original ordering of the measurement tuples as expressed by
the players' preferences over the individual measurement results. In our
simple example it is the change of input state, whilst keeping all other
elements of the physical game unchanged, that ultimately leads to this
possibility of the transformation of the preferences. A general input state
can be expanded in the basis of the measurement so that we have $\left\vert
\psi_{0}\right\rangle =a\left\vert 00\right\rangle +b\left\vert
01\right\rangle +c\left\vert 10\right\rangle +d\left\vert 11\right\rangle $.
It is natural to ask the question as to what are the conditions on the input
state, whilst keeping all other elements unchanged, that will strictly
preserve the original preferences? Before attempting to give a general
perspective on this we consider some special cases.

\subsection{$\left\vert \psi_{0}\right\rangle =a\left\vert 00\right\rangle
+d\left\vert 11\right\rangle $}

Let us generalize the input state given in equation (2) and ask when this
preserves the original preferences. We shall not be fully general in this
approach because we maintain the specific numerical weightings of PD to
express the original preferences, but nevertheless it provides us with an
insight into the way the original game can be transformed with different
inputs. In fact the transformation (or otherwise) of the preferences depends
upon the specific values chosen to represent those preferences. It is possible
to choose numerical weightings that respect the preferences such that those
preferences are also reflected in the matrix of expected outcomes for any
input state, including entangled states. We shall choose, however, the usual
initial preferences of PD expressed by the weightings $\left(  5,3,1,0\right)
$ in order to allow some form of comparison.

With this initial entangled input state and the standard PD weightings and
noting that $\left\vert a\right\vert ^{2}+\left\vert d\right\vert ^{2}=1$ we
find the matrix of expected outcomes

\begin{quote}%
\[%
\begin{tabular}
[c]{c|cc}%
\textit{A}%
$\backslash$%
\textit{B} & $I$ & $F$\\\hline
&  & \\
$I$ & $\left(  1+2\left\vert a\right\vert ^{2},1+2\left\vert a\right\vert
^{2}\right)  $ & $\left(  5-5\left\vert a\right\vert ^{2},5\left\vert
a\right\vert ^{2}\right)  $\\
&  & \\
$F$ & $\left(  5\left\vert a\right\vert ^{2},5-5\left\vert a\right\vert
^{2}\right)  $ & $\left(  3-2\left\vert a\right\vert ^{2},3-2\left\vert
a\right\vert ^{2}\right)  $%
\end{tabular}
\ \
\]
\textbf{Table 3}: \textit{the outcome matrix for the simple 2-player game of
section 2 with the input state} $\left\vert \psi_{0}\right\rangle =a\left\vert
00\right\rangle +d\left\vert 11\right\rangle $
\end{quote}

If we write a general outcome matrix in the following way

\begin{quote}%
\[%
\begin{tabular}
[c]{c|cc}%
\textit{A}%
$\backslash$%
\textit{B} & $I$ & $F$\\\hline
&  & \\
$I$ & $O_{1}$ & $O_{2}$\\
&  & \\
$F$ & $O_{3}$ & $O_{4}$%
\end{tabular}
\ \
\]
\textbf{Table 4}: \textit{general form of the outcome matrix of the pure
strategy 2-player game in which players have 2 choices of action each}.
\end{quote}

then we can see that the game of Prisoner's Dilemma occurs when we have the preferences%

\begin{align}
P_{A}~~~  &  :~~~O_{3}>O_{1}>O_{4}>O_{2}\nonumber\\
& \nonumber\\
P_{B}~~~  &  :~~~O_{2}>O_{1}>O_{4}>O_{3}%
\end{align}
In order to strictly preserve these preferences for our given input state we
therefore require that%

\begin{equation}
5\left\vert a\right\vert ^{2}>1+2\left\vert a\right\vert ^{2}>3-2\left\vert
a\right\vert ^{2}>5-5\left\vert a\right\vert ^{2}%
\end{equation}
which give the conditions under which the preferences of both players are
preserved for this input state. We have plotted the expected outcomes for
Alice as a function of $p=\left\vert a\right\vert ^{2}$ in Figure
\ref{PCGameFigure1} below. We can see that there are 3 regions with each
region giving a different preference ordering for the outcomes.

\begin{figure}[h]
\centerline{\includegraphics[bb=1in 0.4in 9in 9in,scale=0.5]{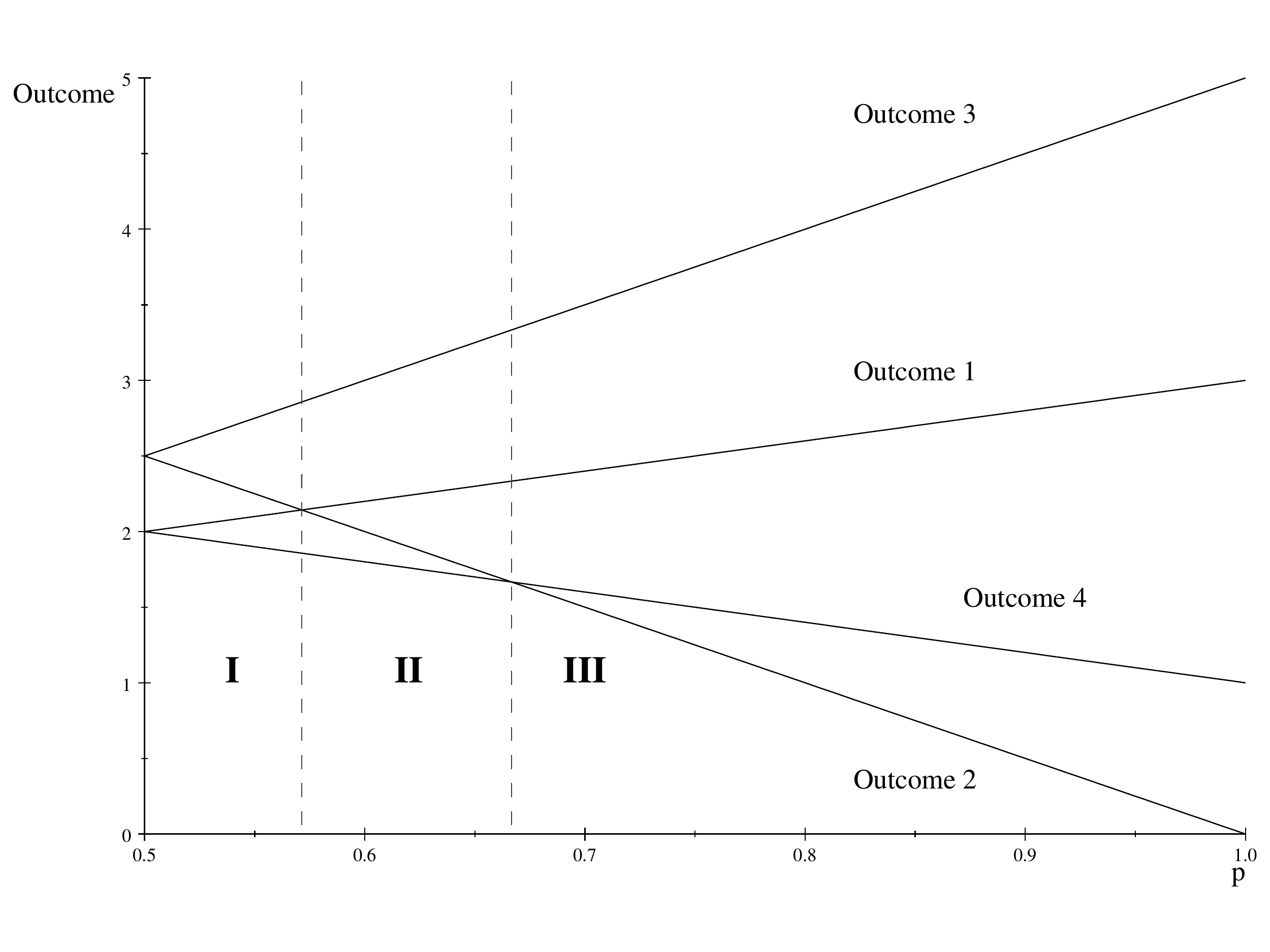}}\caption{
expected outcomes for Alice as a function of $p=\left\vert a\right\vert ^{2}%
$.}%
\label{PCGameFigure1}%
\end{figure}


The different regions can be determined by consideration of the inequalities
in equation (3) and we obtain the regions (where we exclude the boundary points)%

\begin{align}
\text{Region I \ \ }  &  \text{: \ \ }\frac{1}{2}<\left\vert a\right\vert
^{2}<\frac{4}{7}\nonumber\\
& \nonumber\\
\text{Region II \ \ }  &  \text{: \ \ }\frac{4}{7}<\left\vert a\right\vert
^{2}<\frac{2}{3}\nonumber\\
& \nonumber\\
\text{Region III \ \ }  &  \text{: \ \ }\frac{2}{3}<\left\vert a\right\vert
^{2}<1
\end{align}
The preferences for the players for these regions are given in the table below

\begin{quote}%
\[%
\begin{tabular}
[c]{c|cccc}
&  & Alice &  & Bob\\\hline
&  &  &  & \\
$\text{Region I }$ &  & $O_{3}>O_{2}>O_{1}>O_{4}$ &  & $O_{2}>O_{3}%
>O_{1}>O_{4}$\\
&  &  &  & \\
$\text{Region II}$ &  & $O_{3}>O_{1}>O_{2}>O_{4}$ &  & $O_{2}>O_{1}%
>O_{3}>O_{4}$\\
&  &  &  & \\
$\text{Region III }$ &  & $O_{3}>O_{1}>O_{4}>O_{2}$ &  & $O_{2}>O_{1}%
>O_{4}>O_{3}$%
\end{tabular}
\ \
\]
\textbf{Table 5}: \textit{the preferences of the players over the expected
outcomes expressed as preference relations for the different entanglement
regions where the input state is given by} $\left\vert \psi_{0}\right\rangle
=a\left\vert 00\right\rangle +d\left\vert 11\right\rangle $
\end{quote}

In Region III we can see that the players play Prisoner's Dilemma, but in
Region II they play the game of Chicken.

We can see that for $\left\vert a\right\vert ^{2}>\frac{2}{3}$ (and by
symmetry for $\left\vert a\right\vert ^{2}<\frac{1}{3}$ with a switch in the
interpretation of cooperate and defect) the players just play Prisoner's
Dilemma, which might suggest that for these values of the parameter
$\left\vert a\right\vert ^{2}$ the input state is not `quantum' enough to
change the game. We must, however, be careful in making such a claim. Is this
changing of the game by inducing a probability distribution over the
measurements really a non-classical effect? Let us look at the singlet-type
state as input next.

\subsection{$\left\vert \psi_{0}\right\rangle =b\left\vert 01\right\rangle
+c\left\vert 01\right\rangle $}

Once again we maintain the numerical weightings for the original PD game and
note that $\left\vert b\right\vert ^{2}+\left\vert c\right\vert ^{2}=1$. With
this initial state and the available actions of the players the expected
payoff matrix is given by

\begin{quote}%
\[%
\begin{tabular}
[c]{c|cc}%
\textit{A}%
$\backslash$%
\textit{B} & $I$ & $F$\\\hline
&  & \\
$I$ & $\left(  5-5\left\vert b\right\vert ^{2},5\left\vert b\right\vert
^{2}\right)  $ & $\left(  1+2\left\vert b\right\vert ^{2},1+2\left\vert
b\right\vert ^{2}\right)  $\\
&  & \\
$F$ & $\left(  3-2\left\vert b\right\vert ^{2},3-2\left\vert b\right\vert
^{2}\right)  $ & $\left(  5\left\vert b\right\vert ^{2},5-5\left\vert
b\right\vert ^{2}\right)  $%
\end{tabular}
\ \
\]
\textbf{Table 6}: \textit{the outcome matrix for the simple 2-player game of
section 2 with the input state} $\left\vert \psi_{0}\right\rangle =b\left\vert
01\right\rangle +c\left\vert 01\right\rangle $
\end{quote}

\begin{figure}[h]
\centerline{\includegraphics[bb=1in 0.4in 9in 9in,scale=0.5]{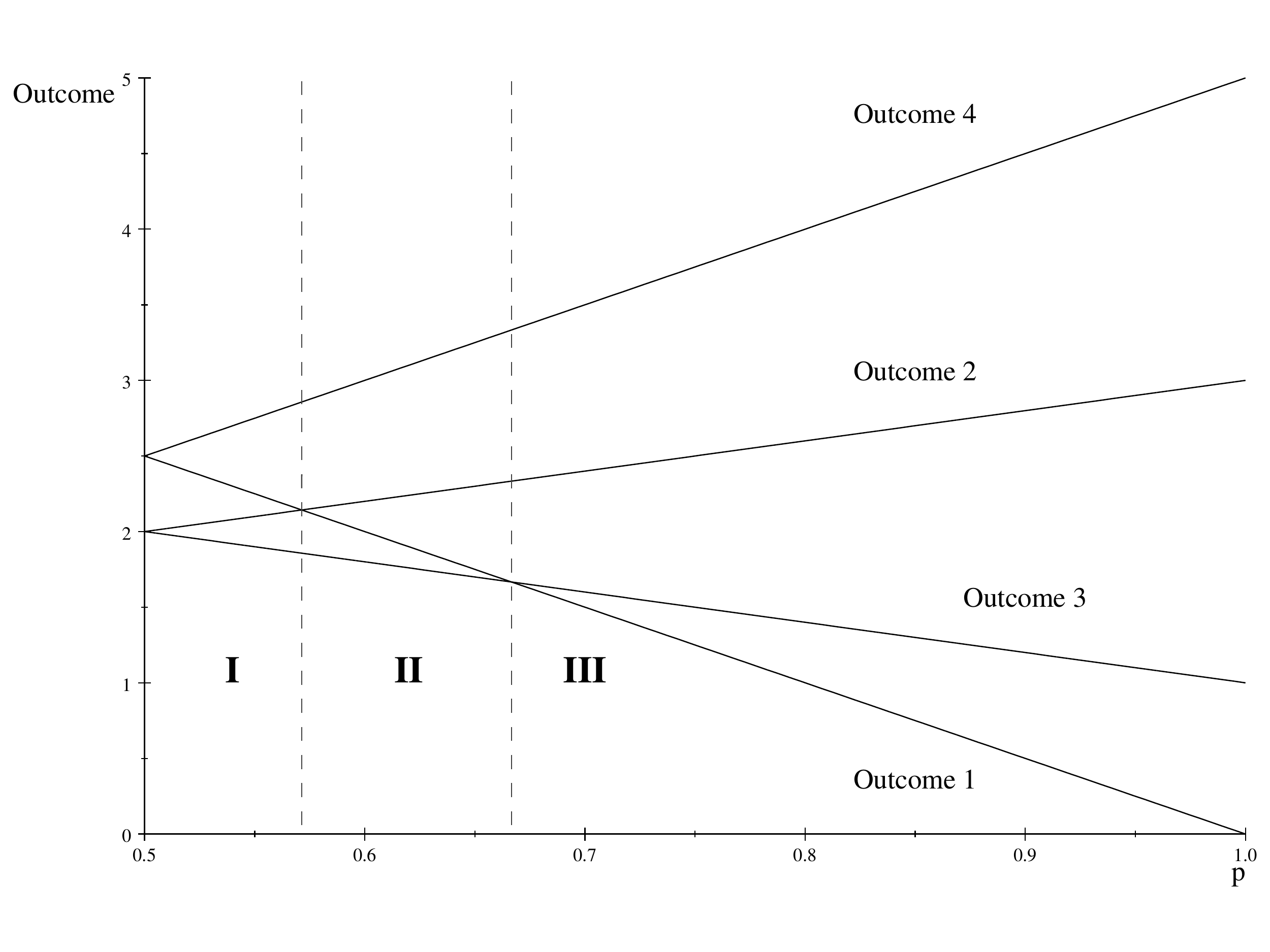}}\caption{
expected outcomes for Alice as a function of $p=\left\vert b\right\vert ^{2}%
$.}%
\label{PCGameFigure2}%
\end{figure}

The expected outcomes for Alice are plotted in Figure \ref{PCGameFigure2}
below as a function of $p=\left\vert b\right\vert ^{2}$

Of course, these are the same lines as before for the input of section 4.1 but
the expected outcomes they represent are different entries in the expected
payoff matrix. We note from Figure \ref{PCGameFigure2} (and analysis of the
conditions for strict preservation of the preferences) that this input state
\textit{always} changes the preference relations from the original with none
of the new preferences over the expected payoff matrix being equivalent to a
game of Prisoner's Dilemma. The regions where the preferences over the
expected outcomes change from one ordering to another are just as before and
the preferences for the players for these regions are given in the table below

\begin{quote}%
\[%
\begin{tabular}
[c]{c|cccc}
&  & Alice &  & Bob\\\hline
&  &  &  & \\
$\text{Region I }$ &  & $O_{4}>O_{1}>O_{2}>O_{3}$ &  & $O_{1}>O_{4}%
>O_{2}>O_{3}$\\
&  &  &  & \\
$\text{Region II}$ &  & $O_{4}>O_{2}>O_{1}>O_{3}$ &  & $O_{1}>O_{2}%
>O_{4}>O_{3}$\\
&  &  &  & \\
$\text{Region III }$ &  & $O_{4}>O_{2}>O_{3}>O_{1}$ &  & $O_{1}>O_{2}%
>O_{3}>O_{4}$%
\end{tabular}
\ \
\]
\textbf{Table 7}: \textit{the preferences of the players over the expected
outcomes expressed as preference relations for the different entanglement
regions where the input state is given by} $\left\vert \psi_{0}\right\rangle
=b\left\vert 01\right\rangle +c\left\vert 01\right\rangle $
\end{quote}

It would not be surprising in quantum PD with this input state, or indeed with
the input state of the previous section with $\frac{1}{2}<\left\vert
a\right\vert ^{2}<\frac{2}{3}$, that the equilibrium payoff might be different
to that of standard PD as we are no longer \textit{actually playing}
Prisoner's Dilemma! The entangled states are often taken to be the most
`non-classical' states possible. Accordingly, it is always tempting to ascribe
any unusual result when working with these states to a `quantum' behaviour.
However, as the analysis of Bell's Theorem shows, pinning down
non-classicality is often surprisingly subtle. In Bell's Theorem for 2
spin-1/2 particles we need to examine correlations between sets of
measurements in different spin directions in order to reveal behaviour that
can be directly attributed to quantum mechanics in the sense that a
`classical' local hidden variable description cannot predict the correct
correlations. Determining what is `quantum' in a quantum game is, in our
opinion, not a trivial issue.

\subsection{A Classical Model}

The feature that the quantum measurement introduces is that the measurement
maps the output state of the players onto an eigenstate of the measurement
with a probability distribution determined by the amplitudes of the
eigenstates in the expansion of the output state in the measurement basis. We
can view this as a noise process. The players are trying to communicate a
particular choice, but noise on the channel gives rise to an error rate. If
the players are aware of the noise and its characteristics then they can build
this knowledge into their strategy. This is exactly what we have in the
quantum situation. So let's model the transmission of the players' choices in
a classical game as a communication over a noisy channel in which the players
are aware of the noise characteristics and can tailor their choices
accordingly. As in the quantum case we will have to deal with a matrix of
expected payoffs which could lead to the playing of a different game by
transformation of the preferences. Can we achieve this transformation of
preferences with such a classical game over a noisy channel?

The simplest case of noise we could consider would be to model the
communication as two independent channels with the same error rate
$\varepsilon$. A simple calculation shows that such a case preserves the
preference relations (or flips them when $\varepsilon>\frac{1}{2}$ but the
original PD is recovered with an interchange of the choices cooperate and
defect). It is of course formally equivalent to the extension to a mixed game
in which the players choose the same probability. A more interesting case
occurs when we consider a \textit{correlated} noise such that both channels,
or neither, experience an error for a given symbol with a rate $\varepsilon$.
Such a correlated noise is, of course, a form of classical noise. We could
imagine the players' signals sent over the same channel and experiencing the
same noise, for example. With this kind of noise, if the players send a pair
of symbols then either both are correct with probability $1-\varepsilon$ or
both are flipped with probability $\varepsilon$.

The expected outcomes for Alice and Bob when they communicate their choices
over such a channel are

\begin{quote}%
\[%
\begin{tabular}
[c]{c|cc}%
\textit{A}%
$\backslash$%
\textit{B} & $I$ & $F$\\\hline
&  & \\
$I$ & $\left(  3-2\varepsilon,3-2\varepsilon\right)  $ & $\left(
5\varepsilon,5-5\varepsilon\right)  $\\
&  & \\
$F$ & $\left(  5-5\varepsilon,5\varepsilon\right)  $ & $\left(  1+2\varepsilon
,1+2\varepsilon\right)  $%
\end{tabular}
\ \
\]
\textbf{Table 8}: \textit{the matrix of expected outcomes for classical PD in
which the players' strategy choices are communicated over a channel with
correlated noise such that both bits, or neither, are flipped.which is just
the same as the expected outcome matrix for the input state }$\left\vert
\psi_{0}\right\rangle =a\left\vert 00\right\rangle +d\left\vert
11\right\rangle $\textit{ considered in section 4.1 for the quantum PD where}
$\epsilon=1-\left\vert a\right\vert ^{2}$.
\end{quote}

So we can see that a game of PD played over channels with this kind of
\textit{correlated} noise will also change the preferences of the players and
in Region II the players will be playing Chicken rather than PD. Thus, there
is nothing particularly quantum mechanical in nature about the transformation
of preferences we obtain for the entangled quantum games considered above.

\subsection{Mixing and Correlated Noise}

Whilst we have not considered the mixed game at all so far, it is instructive
to examine the effect of having a classical correlated noise when we extend a
(classical) game by mixing. As before, we begin with classical PD but now
assume the players will adopt a probabilistic strategy so that Alice chooses
to flip (defect) with probability $p$ and Bob chooses to flip with probability
$q$. As in the previous section they attempt to communicate their choice over
a channel that experiences a correlated noise so that either both bits
representing the players' choice are transmitted error free, or both are
flipped. If we assume the error rate is $\varepsilon$ as before then the joint
probabilities for obtaining the measured results $I$ and $F$ are as follows:%

\begin{align}
P\left(  I,I\right)   &  =\left(  1-p\right)  \left(  1-q\right)  \left(
1-\varepsilon\right)  +pq\varepsilon\nonumber\\
& \nonumber\\
P\left(  I,F\right)   &  =q\left(  1-p\right)  \left(  1-\varepsilon\right)
+p\left(  1-q\right)  \varepsilon\nonumber\\
& \nonumber\\
P\left(  F,I\right)   &  =p\left(  1-q\right)  \left(  1-\varepsilon\right)
+q\left(  1-p\right)  \varepsilon\nonumber\\
& \nonumber\\
P\left(  F,F\right)   &  =pq\left(  1-\varepsilon\right)  +\left(  1-p\right)
\left(  1-q\right)  \varepsilon
\end{align}
The expected outcomes for Alice and Bob now become%

\begin{align}
\left\langle O_{A}\left(  \varepsilon\right)  \right\rangle  &  =\left[
3+2p-3q-pq\right]  \left(  1-\varepsilon\right)  +\left[  1-p+4q-pq\right]
\varepsilon\nonumber\\
& \nonumber\\
\left\langle O_{B}\left(  \varepsilon\right)  \right\rangle  &  =\left[
3+2q-3p-pq\right]  \left(  1-\varepsilon\right)  +\left[  1-q+4p-pq\right]
\varepsilon
\end{align}
The $\left(  1-\varepsilon\right)  $ part of this expected outcome is just the
usual expected outcome from the mixed PD without any noise. In this noiseless
case the players are forced to the equilibrium position $\left(  F,F\right)  $
just as the non-mixed game and the expected outcome is $\left(  1,1\right)  $.
However, the noise term now changes this expected outcome. If $\varepsilon=1$
then the players would play the equilibrium position $\left(  I,I\right)  $
with an expected outcome of $\left(  1,1\right)  $. The actual choice of
probability the players make is a function of the error rate $\varepsilon$ and
we can see that their best response is given by the choice $p=q=1-\varepsilon
$. This yields the expected outcomes for the players%

\begin{equation}
\left\langle O_{A}\left(  \varepsilon\right)  \right\rangle =\left\langle
O_{B}\left(  \varepsilon\right)  \right\rangle =1+5\varepsilon-5\varepsilon
^{2}%
\end{equation}

\begin{figure}[h]
\centerline{\includegraphics[bb=1in 0.4in 9in 9in,scale=0.5]{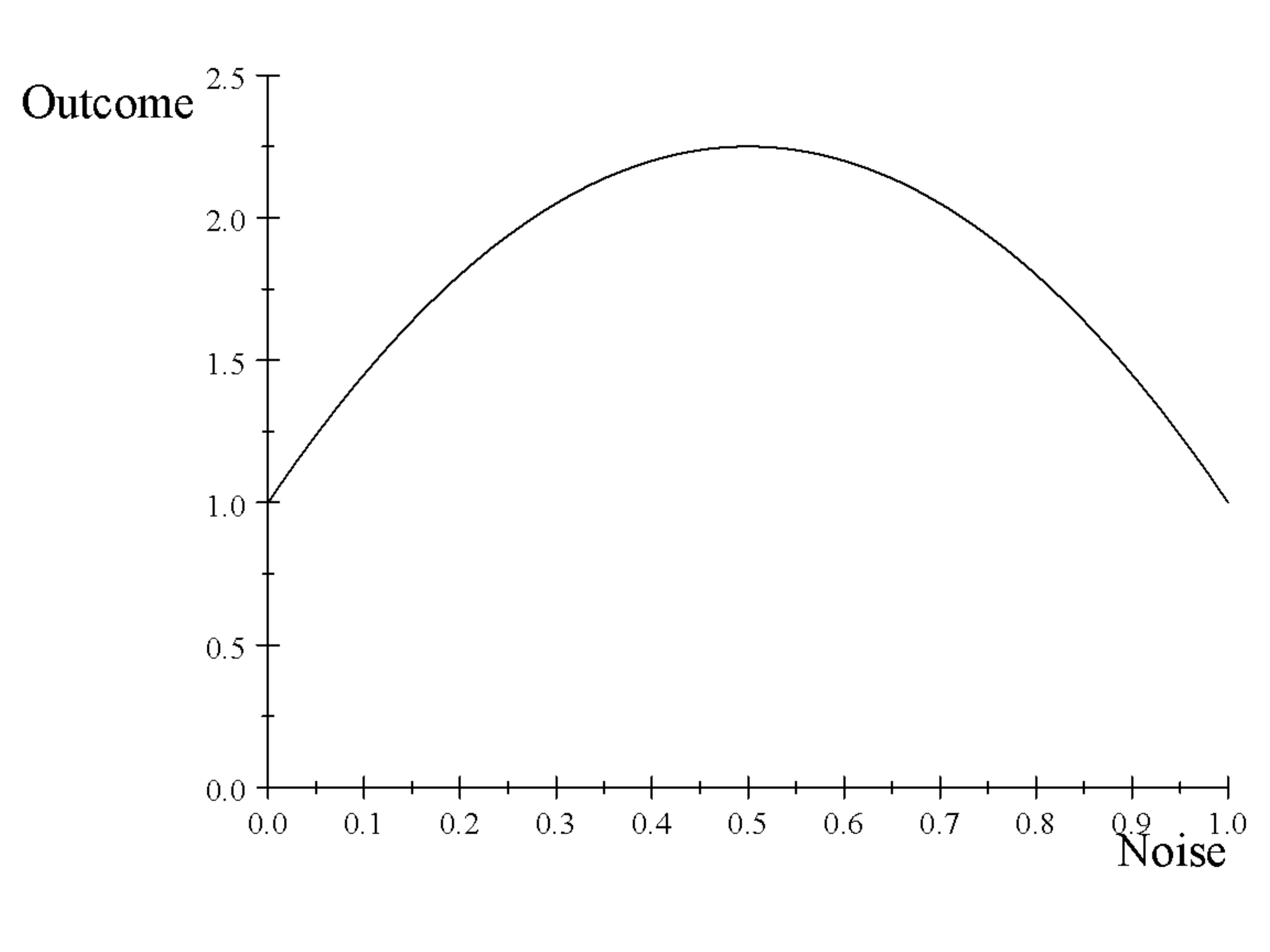}}\caption{
Expected outcomes for Alice as a function of the error rate $\epsilon$.}%
\label{PCGameFigure3}%
\end{figure}
This expected outcome is plotted above in Figure \ref{PCGameFigure3} and we
can see that the maximum value is obtained when $\varepsilon=\frac{1}{2}$ and
this gives an expected outcome for the players of $\frac{9}{4}$ which is an
improvement on their equilibrium output in the noise-free case (or the
all-noise case). This is, of course, nothing more than a uniform distribution
of the possible outcomes and the same result is obtained for quantum games in
which there is maximal decoherence [19,20]. In this case, it is the
correlated, but classical, noise that is giving an enhanced equilibrium payoff
for the players in all regions except the boundary points (noise-free or all-noise).

\subsection{Noise in the Quantum Game}

Now let us consider the case where the players play their version of
Prisoner's Dilemma with the input state $\left\vert \psi_{0}\right\rangle
=a\left\vert 00\right\rangle +d\left\vert 11\right\rangle $. We shall consider
that they attempt to communicate their choice by sending their respective
particles over some channel to be measured. Thus we now have a quantum
channel. We shall suppose that there is some noise source on this channel. In
this case we shall not be too concerned with the details of the noise, but
merely suppose that it is sufficient to rapidly suppress the off diagonal
coherences in the density matrix. Such a suppression of off-diagonal
coherences is, of course, a general feature of open quantum systems [26-28].
If we restrict the available operations, as before, to this binary choice of
whether to flip or not in the measurement basis, then after this decohering
noise process the density matrix description of the state that arrives at the
measurement apparatus is given, for each of the possible choices of the
players, as%

\begin{align}
\rho_{II}  &  =\left\vert a\right\vert ^{2}\left\vert 00\right\rangle
\left\langle 00\right\vert +\left\vert d\right\vert ^{2}\left\vert
11\right\rangle \left\langle 11\right\vert \nonumber\\
& \nonumber\\
\rho_{IF}  &  =\left\vert a\right\vert ^{2}\left\vert 01\right\rangle
\left\langle 01\right\vert +\left\vert d\right\vert ^{2}\left\vert
10\right\rangle \left\langle 10\right\vert \nonumber\\
& \nonumber\\
\rho_{II}  &  =\left\vert a\right\vert ^{2}\left\vert 10\right\rangle
\left\langle 10\right\vert +\left\vert d\right\vert ^{2}\left\vert
01\right\rangle \left\langle 01\right\vert \nonumber\\
& \nonumber\\
\rho_{II}  &  =\left\vert a\right\vert ^{2}\left\vert 11\right\rangle
\left\langle 11\right\vert +\left\vert d\right\vert ^{2}\left\vert
00\right\rangle \left\langle 00\right\vert
\end{align}
The expected outcomes for the players are given by the expected outcome matrix

\begin{quote}%
\[%
\begin{tabular}
[c]{c|cc}%
\textit{A}%
$\backslash$%
\textit{B} & $I$ & $F$\\\hline
&  & \\
$I$ & $\left(  1+2\left\vert a\right\vert ^{2},1+2\left\vert a\right\vert
^{2}\right)  $ & $\left(  5-5\left\vert a\right\vert ^{2},5\left\vert
a\right\vert ^{2}\right)  $\\
&  & \\
$F$ & $\left(  5\left\vert a\right\vert ^{2},5-5\left\vert a\right\vert
^{2}\right)  $ & $\left(  3-2\left\vert a\right\vert ^{2},3-2\left\vert
a\right\vert ^{2}\right)  $%
\end{tabular}
\ \ \ \
\]
\textbf{Table 9}: \textit{the matrix of expected outcomes for the 2-player
game with input state }$\left\vert \psi_{0}\right\rangle =a\left\vert
00\right\rangle +d\left\vert 11\right\rangle $\textit{ in which the players
particles are sent over a noisy quantum channel that leads to suppression of
the off-diagonal coherences.}
\end{quote}

Which is precisely the same as that for the game played in the noiseless case
considered in section (4.1). In other words, the off-diagonal components in
the density matrix in the noiseless case are not contributing to the
determination of the expected outcomes. This is only to be expected since any
single pair of measurements on the separate systems can, at most, only access
half the information contained within the quantum correlation [29,30]. If we
expressed our entangled state in the Schmidt basis, and made measurements of
the Schmidt observables (this is considered within the context of quantum
games in [10,11]), then we would access precisely half the information
contained within the quantum correlation [29,30]. In other words, the enhanced
equilibrium obtained for quantum games in which a single measurement is made
in each the subspaces is a result of a \textit{classical} correlation because
the off-diagonal interference terms are not being accessed in such a measurement.

We note that a similar problem has been studied in more generality by
Shimamura et. al. [18] in which they consider the difference between an
entangled state input and its classical counterpart in games of the EWL type
(see also Chen et. al. [19] who consider a decoherence protocol for Quantum
Prisoner's Dilemma). Both [18] and [19] are different to the situation we
envisage here in which the decoherence occurs during the transmission of the
quantum states to the measurement device (or referee). In [18] the referee
employs the disentangling transformation before measurement and the full space
of local unitary operations is allowed by each player. The formal
correspondence discussed in section 3 above is no longer applicable because
the decoherence destroys the symmetry and the placement of the decohering
process in the chain of events becomes significant. Furthermore in both [18]
and [19] the referee performs a disentangling operation before measurement,
which amounts to a re-entanglement in the case of a separable mixed state input.

Our purpose here is not to examine the effect of decoherence in general (see
for example, [20]) but to provide another illustration that, for simple
2-player games of the form considere here, it is correlation, and not quantum
correlation, that is the interesting feature. Of course a more general
treatment is required to determine the game types and conditions under which
quantum correlations do become significant. We shall examine the ramifications
of this, and the results of the previous sections, for the interpretation of
games such as EWL [1] and MW [21] elsewhere.

\section{A General Approach}

In the preceding sections we examined some very simple, and restricted,
quantum games in order to gain some insight into the role of correlation in
quantum games. An obvious question is whether the results obtained depend in
some way on the nature of the restriction imposed (the choices cooperate or
defect being the only operations available to the players).

Consider a game played with quantum mechanical objects and operations that
obey the laws of quantum mechanics. The game consists of the following [23-25]:

\begin{itemize}
\item An input state $\left\vert \psi_{0}\right\rangle $\ that is assumed to
be known by the players

\item Actions available to player \textit{A} described by a finite set of
unitary operators $\left\{  \hat{\alpha}_{1},\ldots,\hat{\alpha}_{i}%
,\ldots,\hat{\alpha}_{n}\right\}  $

\item Actions available to player \textit{B} described by a finite set of
unitary operators $\left\{  \hat{\beta}_{1},\ldots,\hat{\beta}_{j},\ldots
,\hat{\beta}_{m}\right\}  $

\item The actions of the players on the input state produce some output state
$\left\vert \psi_{ij}\right\rangle $ that is characterized by the choice of
$\hat{\alpha}_{i}$ and $\hat{\beta}_{j}$ by players \textit{A} and \textit{B},
respectively. There are $n\times m$ possible output states from this game for
a given input state $\left\vert \psi_{0}\right\rangle $

\item A projective measurement $\hat{M}$ on the output state that produces an
eigenstate $\left\vert m_{i}\right\rangle $\ of the measurement operator where
there are $r$ such eigenstates. We assume non-degenerate eigenvalues so that
each measurement result can be unambiguously identified with a measurement eigenstate.

\item The players each have a different preference relation over the
measurement eigenstates. Accordingly, we shall use the terms
\textit{preference basis} and \textit{measurement basis} interchangeably. The
preference relations therefore induce a preference relation for each player
over the set of possible output states $\left\vert \psi_{ij}\right\rangle $

\item We shall encapsulate the notion of preference by assigning a numerical
value to each measurement eigenstate for each player such that a higher
numerical value indicates a greater preference for that player.

\item We shall assume that (nominally) each player has some object upon which
to act so that the Hilbert space is described by $H=H_{A}\otimes H_{B}$. Note
that this does \textit{not}, therefore, imply that the unitary operations
available to the players act \textit{only} in their respective subspaces
\end{itemize}

The players, as noted above, have some preference over the measurement
eigenstates so that the output state produced can be expressed in this
measurement, or preference, basis as follows%
\begin{equation}
\left\vert \psi_{ij}\right\rangle =\sum_{i=1}^{r}\left\langle m_{i}\mid
\psi_{ij}\right\rangle ~\left\vert m_{i}\right\rangle
\end{equation}

Upon measurement, the result $\left\vert m_{l}\right\rangle $ is mapped to a
numerical value in accordance with the preference relations as%

\begin{equation}
\left(  \left\vert m_{l}\right\rangle \right)  \longrightarrow\left(
\omega_{l}^{A},\omega_{l}^{B}\right)
\end{equation}
where $\omega_{l}^{A}$ is the outcome for player \textit{A} if the result of
the measurement yields the eigenvalue $\left\vert m_{l}\right\rangle $. We can
formally combine the measurement and assignment of outcomes into the single
Hermitian `outcome' operators%

\begin{align}
\hat{\omega}_{A}  &  =%
{\displaystyle\sum\limits_{i=1}^{r}}
\omega_{i}^{A}\left\vert m_{i}\right\rangle \left\langle m_{i}\right\vert
\nonumber\\
& \nonumber\\
\hat{\omega}_{B}  &  =%
{\displaystyle\sum\limits_{i=1}^{r}}
\omega_{k_{B}}^{B}\left\vert m_{i}\right\rangle \left\langle m_{i}\right\vert
\end{align}
In general, the output state will be a superposition over the preference bases
and will not be an eigenstate of the measurement operator (or equivalently the
outcome operators). There will therefore be a distribution over the output
tuples for any given choice of $\hat{\alpha}_{j}$ and $\hat{\beta}_{k}$. Thus
for each choice of $\hat{\alpha}_{j}$ and $\hat{\beta}_{k}$ there will be an
\textit{average} outcome tuple expressed as the expected value of the outcome operators%

\begin{equation}
\left(  \left\langle \hat{\omega}_{A}\right\rangle _{jk},\left\langle
\hat{\omega}_{B}\right\rangle _{jk}\right)  =\left(
{\displaystyle\sum\limits_{i=1}^{r}}
\omega_{i}^{A}\left\vert \left\langle m_{i}\mid\psi_{jk}\right\rangle
\right\vert ^{2}~,~%
{\displaystyle\sum\limits_{i=1}^{r}}
\omega_{i}^{B}\left\vert \left\langle m_{i}\mid\psi_{jk}\right\rangle
\right\vert ^{2}\right)
\end{equation}
Each expected outcome tuple can therefore be thought of as an entry in an
$n\times m$ matrix of outcome tuples, just as we would describe any 2-player
game. Thus the quantum mechanical game is entirely equivalent, as far as the
players are concerned, to a classical game in which the outcomes relating to
the choices of the players are described by this matrix. The fact that these
outcomes are derived from a quantum measurement and the resultant
probabilities is utterly irrelevant. The players play the game
\textit{according to the outcomes expressed in this expected payoff matrix}.
The game is defined not by their original preferences over the measurement
results, but by the induced preferences as expressed by the matrix of expected
outcomes. In effect, the quantum mechanical measurement has the potential to
change the players' preferences to those expressed in the expected payoff
matrix. So although the players start off with a set of preferences over the
results of the measurement they act \textit{as if} they had a new set of
preferences, given by the expected outcomes. The players choose their
strategies according to this new matrix. \textit{It is this matrix which
defines the actual game they are playing}.

The expectation values defined in equation (18) are those considered by Cheon
and Tsutsui [17], in which they show that each $\left\langle \hat{\omega}%
_{A}\right\rangle _{jk}$ and $\left\langle \hat{\omega}_{B}\right\rangle
_{jk}$ can be considered to arise from a pseudo-classical part and quantum
interference terms. So each element in our matrix of expected payoffs can be
thought to arise partly from some quantum interference term in which a
different quantum interference term is obtained for each choice of operation
(strategy) by Alice and Bob. So whilst each separate entry into the payoff
matrix may be thought of in this manner, the entire matrix is just a set of
classical probabilities that can, as we argue below, be reproduced by
modelling the game classically as a communication channel in which we allow
the possibility of classical correlated noise. Once again, we emphasize that
it is the matrix of expected payoffs that \textit{defines} the game the
players \textit{actually} play, and not whatever complicated physical
mechanism we have used to produce this matrix.

It is at this point we must ask what is quantum mechanical about games of this
type? It is irrelevant to the players whether the expected payoff matrix that
defines the game they are playing has been generated by some quantum process,
or whether the entries in the matrix are assigned to measurement of classical
coins, just as in any standard classical 2-player game. There is nothing
particularly quantum-mechanical about a matrix of tuples and any quantum game
of this form can be implemented entirely by classical objects with a given
functional mapping of measurement to outcomes. The specific functional
decomposition of the game that has generated the final game function is
irrelevant to the players; the quantity they analyze is the matrix of expected
outcomes. We believe that van Enk and Pike [3] were right to be uncomfortable
about the `quantum' claim for games of this type.

A matrix of expected outcome tuples is generated in any pure strategy game,
quantum or classical, where there is some probability distribution over the
measurement results. In the quantum case the measurement induces this
distribution, but we can similarly imagine a classical game in which the
measurement process is imperfect. The example of the implementation of a
classical game as a communication over a noisy channel considered above is
just one way of realizing such a distribution of measurement results in the
classical case. For a classical game with a distribution over $r$ possible
measurement results, $m_{i}$, we have the conditional distribution $P\left(
m_{i}\mid\alpha_{j},\beta_{k}\right)  $ which gives the probability of
obtaining the measurement result $m_{i}$ given that the players choose the
operations $\alpha_{j}$ and $\beta_{k}$. In this case the expected outcomes
for Alice are%
\begin{equation}
\left\langle \omega_{A}\right\rangle _{jk}=\sum_{i=1}^{r}P\left(  m_{i}%
\mid\alpha_{j},\beta_{k}\right)  \omega_{i}^{A}%
\end{equation}
with a similar expression for Bob's outcomes. By modelling the classical game
as a communication of a binary number representing the strategy choice of the
players we can see that a classical noise process on the channel such that the
channel transition probabilities are%
\begin{equation}
P\left(  m_{i}\mid\alpha_{j},\beta_{k}\right)  =\left\vert \left\langle
m_{i}\mid\psi_{jk}\right\rangle \right\vert ^{2}%
\end{equation}
will reproduce the results of the quantum game. Thus by assuming a channel
with a classical correlated noise we can reproduce the results of a quantum
entangled game where we assume a single projective measurement is performed on
the resultant output state. The game, classical or quantum, can be thought of
as a communication channel where the input symbols are the $n\times m$
strategy choices $\alpha_{j}\beta_{k}$ and the $r$ output symbols are the
measurement results $m_{i}$. We are free to model the noise on such a
classical channel with any legitimate set of conditional probabilities and
these probabilities represent a classical noise (although we may need some
peculiar classical noise process to generate the results of a particular
quantum game, it is still classical).

The modelling of a pure strategy game as a communication channel in which the
input symbol is chosen according to preferences over the output symbols is
instructive. For convenience we shall assume the players have strategy sets of
equal size where $n=m=2^{\mu}$ so that a classical game can be represented as
a channel over which the players each communicate an $\mu$-bit binary string.
We can implement this classical game using spin-1/2 particles prepared in the
state $\left\vert 00\ldots0\right\rangle $ in some spin basis where the
players can perform a flip or a no-flip operation in this basis and the
measurement is performed on each particle in this basis. This can be
considered to be an expensive quantum implementation of the classical game.
The expensive quantum implementation can also be thought of as a quantum
communication channel over which qubits are transmitted.

The expensive quantum implementation can now be altered in 3 obvious ways:

\begin{itemize}
\item the initial state is prepared as $\left\vert \bar{0}\bar{0}\ldots\bar
{0}\right\rangle $ in some other basis

\item the initial state is $\left\vert 00\ldots0\right\rangle $ but the
players are given flip and no-flip in some basis aligned at some angle to the
basis of the input states

\item the initial state is $\left\vert 00\ldots0\right\rangle $ and the
players are given flip and no-flip in this basis, but the measurement of the
qubits is now performed in some other basis aligned at some angle to the input basis
\end{itemize}

Of course we can imagine any combination of these things, or consider
different bases for each qubit, for example. The point here is that we have
changed the quantum implementation so that the output state produced by the
players is no longer an eigenstate of the measurement operator and this
induces a distribution over the measurement results, which in turn induces
preferences over the expected outcomes that can lead to a different game form
than that described by the preferences over the individual measurement
results. In effect, the measurement induces \textit{noise} on the quantum
communication channel.

This changing of the quantum implementation of the game can be thought of as a
kind of game extension in which the original game pertains when we adjust the
alignments to yield the zero noise case. In order to draw a sensible
quantum/classical comparison, therefore, one must compare the noisy quantum
channel with a noisy classical channel in the context of the application to
the description of a game. It is clear that in the unentangled case the
expensive quantum implementation can be modelled as a classical game where
$2\mu$ bits replace the $2\mu$ qubits such that the noise characteristics of
the classical channel reproduce the measurement-induced noise characteristics
of the quantum channel.

If we now extend the quantum implementation to allow entanglement, the above
arguments show that, for the situation where a projective measurement is made
on the output state in the quantum case, the game can be modelled as a
classical communication channel in which we allow the possibility of
correlated noise. In both cases the players transmit $2\mu$ bits or qubits
over the channel. The single projective measurement is not sensitive enough to
distinguish between classical and quantum correlations in these 2-player pure
strategy finite games. This is essentially for the same reason that we require
more than just a single joint probability distribution to distinguish between
hidden variable models and quantum mechanics in tests of Bell's inequality. In
tests of local realism we need to establish the non-existence of the joint
distribution $P\left(  A,B,C\right)  $ that correctly reproduces the marginal
distributions $P\left(  A,B\right)  $ and $P\left(  A,C\right)  $, for
example. Such a distribution only exists if the marginals satisfy the Bell
inequality. In a sense, the classical communication over a noisy channel can
be thought of as a hidden variable implementation of the quantum game and from
this perspective it is not surprising that a 2-player quantum game of the form
we have considered does not display non-classical behaviour.

Other authors have considered more general formulations of quantum games. Of
particular note is the work of Lee and Johnson [16] who show that finite
classical games are a strict subset of quantum games, as we would expect. In
the context of communcation channels this is expressing the fact that any
classical communication channel with a finite input alphabet is a subset of
the possible quantum channels. It is how these channels are exploited that
determines whether they display quantum or classical characteristics. Lee and
Johnson show that a given classical game can have a more efficient
implementation using quantum objects in terms of the relative number of bits
and qubits, respectively. This is reminiscent of the ability of quantum
channels to transmit classical information using fewer qubits [31]. Here we
are interested in a converse (and more restricted) question ; whether a
quantum game can be modelled by a game played with classical objects in the
context of the 2-player pure games with finite strategy sets in which a single
projective measurement is made on the output. We believe that our analysis
offers an insight into where we need to look for game properties that display
necessarily quantum-mechanical features.

The more general form of the 2-player game we have considered assumes a finite
strategy set for the players, and that a single projective measurement is made
on the resultant output state. Furthermore we have assumed a \textit{pure}
strategy game. The simple example of the classical mixed game with correlated
noise considered above shows that similar considerations may also apply in the
more general mixed game case. We consider such situations elsewhere.

\section{Discussion}

There is no doubt that the pioneering work [1,2] that brought together game
theory and quantum mechanics represented a new and original direction in both
fields. There has been much work since on various quantum game scenarios,
usually focusing on the use of entangled states in games [1-21] (we have
referenced only a very small selection of the work that has been done). In our
previous work [23-25] we argued that a game should be seen as something that
can actually be played. In other words there is a physical implementation of a
game with real objects that obey the laws of physics. With this perspective
the necessary elements required to actually play a game can be identified.
These elements are; preparation of some initial state, operations by the
players on that state, a resulting output state that is subject to a
measurement, and a mapping of the results of that measurement to given
outcomes. In order to call such a thing a game, rather than just an experiment
in physics, we require that the players have some preference over the
outcomes, which ultimately determines which operation on the input state they
will choose.

In the games we've discussed here we have assumed that a single measurement on
the output state produced by the players is performed. This measurement
remains fixed however many times we play the game. The simple restricted
examples examined consider a measurement of spin in the $z$-direction for
Alice's particle and a measurement of spin, also in the $z$-direction, for
Bob's particle. Experimentally, therefore, we are only accessing information
about the correlations between these two specific observables. This
measurement cannot uncover the full richness of the quantum correlations
inherent in an entangled state of 2 particles; it is only accessing
\textit{some} of the information about the quantum correlations [29,30]. The
specific examples of games considered are only allowing us to probe
correlations between the spin-$z$ and spin-$z$ measurements, and that can't
give us enough information to decide whether it's quantum or classical
behaviour we're seeing. The information that can be recovered about the
correlations from this kind of fixed measurement is not sufficient to
distinguish between the classical and quantum nature of the correlation. The
more general analysis for 2-player pure strategy games in which the players
can choose from finite strategy sets shows that a single measurement of the
output is also insufficient to access the quantum nature of the correlation.
In order to do that we need to compare correlations for multi-player games or
in 2-player quantum games in which different measurement angles are selected,
for example, just as we need to in establishing the experimental violation of
Bell's inequality [6-9].

The general analysis of these 2-player pure strategy games shows that the
\textit{same} game can be played either with quantum objects in which a single
measurement is made, or as a classical communication channel in which the
players know the noise characteristics. Where correlation features in the
quantum game, when entanglement is introduced, the classical version of the
game as a communication channel requires a classical correlated noise. The
important point to note is that the single measurement in the quantum case
reduces everything to a set of probabilities that can be achieved by an
\textit{equivalent} classical communication channel. We do not require quantum
objects to play games of this form. In other words, given a set of operations
and an associated payoff matrix, there is no way for the players to determine
whether they are playing with quantum or classical objects in these games.

The reason we can reproduce these general 2-player pure strategy quantum games
as a classical communication channel is that we assume a single (joint)
measurement that produces the probability distribution over the results that
are then mapped to the expected outcomes via some payoff function. With a
single joint measurement the probabilities can always be reproduced by a
classical system. Overall, the game is a function that takes some inputs and
performs a computation on those inputs. If that computation can be achieved by
classical systems then it seems to us that the underlying game is essentially
classical even if implemented by quantum objects.

In mathematical terms, then, a 2-player game is nothing more than a function
that takes a pair of inputs representing the choices of the players and maps
these to outcomes. The physical elements required to play a game are nothing
more than a particular functional decomposition of this overall game function.
In these terms, therefore, we can see that for the initial example based on
the MW protocol [21] the function that represents the game of Chicken can be
decomposed into functional elements that look like a version of quantum
Prisoner's Dilemma, or it can be functionally decomposed as a game played over
a classical communication channel with correlated noise. In general,
therefore, if we are to observe genuine quantum behaviour in a game we must
consider richer game structures that allow us to probe the quantum regime of
the correlation, and effectively perform a quantum computation on the inputs
in order to produce the outcomes. Games that allow us to do this are
multi-player games [9,32] or games in which the final outcomes are determined
from comparison of the results of a sequence of games in which different
meaurements are made.

\bigskip

\textbf{References}

\begin{enumerate}
\item J. Eisert, M. Wilkens and M. Lewenstein, \textit{Quantum Games and
Quantum Strategies}, Phys. Rev. Lett., \textbf{83}, 3077-3080, (1999)

\item D. Meyer, \textit{Quantum Strategies, }Phys. Rev. Lett., \textbf{82},
1052-1055, (1999)

\item S. J. van Enk and R. Pike, \textit{Classical Rules in Quantum Games},
Phys. Rev. A, \textbf{66,} 024306, (2002)

\item S. A. Bleiler, \textit{A Formalism for Quantum Games I - Quantizing
Mixtures}, Portland State University, preprint, available at
http://arxiv.org/abs/0808.1389, (2008)

\item J. Shimamura, \c{S}. K. \"{O}zdemir, F. Morikoshi and N. Imoto,
\textit{Entangled states that cannot reproduce original classical games in
their quantum version}, Phys. Lett \ A, \textbf{328}, 20-25, (2004)

\item A. Iqbal and D. Abbott, \textit{Constructing quantum games from a system
of Bell's inequalities}, Phys. Lett. A, \textbf{374}/31-32, 3155-3163, (2010)

\item A. Iqbal\textit{, Playing games with EPR-type experiments}, J. Phys. A,
\textbf{38}/43, 9551-9564, (2005)

\item A. Iqbal and S. Weigert, \textit{Quantum correlation games}, J. Phys. A,
\textbf{37}, 5873-5885, (2004)

\item C.D. Hill, A.P. Flitney and N.C. Menicucci, \textit{A competitive game
whose maximal Nash-equilibrium payoff requires quantum resources for its
achievement}, Phys. Lett. A \textbf{374, }3619 (2010)

\item T. Ichikawa, T. Cheon and I. Tsutsui, \textit{Quantum Game Theory Based
on the Schmidt Decomposition}, J.Phys.A, \textbf{41}, 135303, (2008)

\item T. Ichikawa and I. Tsutsui, \textit{Duality, Phase Structures and
Dilemmas in Symmetric Quantum Games}, Ann. Phys., \textbf{322}, 531-551, (2007)

\item R. Cleve, P. Hoyer, B. Toner and J. Watrous, \textit{Consequences and
Limits of Nonlocal Strategies}, preprint, available at http://arxiv.org/abs/quant-ph/0404076v2

\item R. Renner and S. Wolf, \textit{Towards Characterizing the Non-Locality
of Entangled Quantum States}, preprint, available at
http://arxiv.org/abs/quant-ph/0211019, (2002)

\item S. E. Landsburg, \textit{Nash Equilibria in Quantum Games}, Proc. Am.
Math Soc., \textbf{139}, 4423-4434, (2011)

\item G. Dahl and S. Landsburg, \textit{Quantum Strategies in Noncooperative
Games}, preprint, available at http://www.landsburg.com/dahlcurrent.pdf

\item C. F. Lee and N. Johnson, \textit{Efficiency and Formalism of Quantum
Games}, Phys. Rev. A \textbf{67}, 022311 (2003)

\item T. Cheon and I. Tsutsui, \textit{Classical and Quantum Contents of
Solvable Game Theory on Hilbert Space}, Physics Letters A, \ \textbf{348}
147-152 (2006)

\item J. Shimamura, S. \"{O}zdemir, F. Morikoshi, and N. Imoto,
\textit{Quantum and Classical Correlations Between Players in Game Theory},
International Journal of Quantum Information, \textbf{2}, 79--89 (2004)

\item L.K. Chen, H. Ang, D. Kianga, L.C. Kwek, and C.F. Loc, \textit{Quantum
Prisoner Dilemma under Decoherence}, Physics Letters A, \textbf{316}, 317--323 (2003)

\item A. P. Flitney, and D. Abbott, \textit{Quantum Games with Decoherence},
Journal of Physics A, \textbf{38}, 449, (2005)

\item L. Marinatto and T. Weber, \textit{A quantum approach to static games of
complete information}, Phys. Lett. A, \textbf{272}, 291-303, (2000)

\item K. Binmore, \textit{Playing for Real: A Text on Game Theory}, OUP USA, (2007)

\item S. J. D. Phoenix and F. S. Khan, \textit{Playing Games with Quantum
Mechanics}, preprint available at http://arxiv.org/abs/1202.4708, (2012)

\item F. S. Khan and S. J. D. Phoenix, \textit{Nash equilibrium in quantum
superpositions}, Proceedings of SPIE, \textbf{8057,} 80570K-1, (2011)

\item F. S. Khan and S. J. D. Phoenix, \textit{Gaming the Quantum}, Quant.
Inf. Comp., \textbf{13}, 0231-0244, (2013)

\item W. H. Zurek, \textit{Pointer Basis of Quantum Apparatus: Into what
Mixture does the Wave Packet Collapse}?, Phys. Rev D, \textbf{24}, 1516--1525, (1981)

\item W. H. Zurek, \textit{Environment-Induced Superselection Rules}, Phys.
Rev. D, \textbf{26}, 1862--1880, (1982)

\item W. H. Zurek, \textit{Decoherence, einselection, and the quantum origins
of the classical}, Rev. Mod. Phys., \textbf{75}, 715, (2003) (available at http://arxiv.org/abs/quant-ph/0105127)

\item S. M. Barnett and S. J. D. Phoenix,\textit{\ Information Theory,
Squeezing and Quantum Correlations}, Phys. Rev. A, \textbf{44}, 535, (1991)

\item S. M. Barnett and S. J. D. Phoenix, \textit{Information-Theoretic Limits
to Quantum Cryptography}, Phys. Rev. A, \textbf{48,} R5, (1993)

\item S.M. Barnett, R. Loudon, D.T. Pegg, and S.J.D. Phoenix,
\textit{Communication Using Quantum States}, J. Mod. Opt, \textbf{41},
2351-2373 (1994)

\item A.P. Flitney and D. Abbott, \textit{Multiplayer quantum minority game
with decoherence}, Quant. Inf. Comput, \textbf{7,} 111, (2007)
\end{enumerate}

\end{document}